\input harvmac

\Title{\vbox{\baselineskip12pt\hbox{KUCP-0166}\hbox{{\tt hep-th/0008216}}}}
{\vbox{\centerline{Gravitational Stability and Screening Effect}\vskip12pt\centerline{
from $D$ Extra Timelike Dimensions}}}

\centerline{Satoshi Matsuda\footnote{$^\dagger$}{matsuda@phys.h.kyoto-u.ac.jp}}
{\it \centerline{Department of Fundamental Sciences, FIHS}
\centerline{Kyoto University, Kyoto 606-8501, Japan}}
\medskip
\centerline{Shigenori Seki\footnote{$^\ddagger$}{seki@phys.h.kyoto-u.ac.jp} }
{\it \centerline{Graduate School of Human and Environmental Studies}
\centerline{Kyoto University, Kyoto 606-8501, Japan}}

\vskip .3in

\centerline{{\bf abstract}}
We study $(3+1)+D$ dimensional spacetime, where $D$ extra dimensions are 
timelike. Compactification of the $D$ timelike dimensions leads to tachyonic 
Kaluza-Klein gravitons. We calculate the gravitational self-energies of massive spherical bodies 
due to the tachyonic exchange, discuss their stability, and 
find that the gravitational force is screened 
in a certain number of the extra dimensions. 
We also derive the exact relationship between the Newton constants in the full $4+D$ dimensional 
spacetime 
with the $D$ extra times and the ordinary Newton constant of our 4 dimensional world. 

\Date{August 2000}

\newsec{Introduction}
Extra dimensions have been studied in a variety of contexts. Especially in
recent works 
\ref\ADD{I.Antoniadis, S.Dimopoulos and G.Dvali, 
``Millimeter-range forces in superstring theories with weak-scale compactification'', 
Nucl. Phys. {\bf B516} (1998) 70, {\tt hep-ph/9710204}.}\nref\HDD
{N.Arkani-Hamed, S.Dimopoulos and G.Dvali, ``The hierarchy problem and new dimensions 
at a millimeter'', Phys. Lett. B {\bf 429} (1998) 263,  
{\tt hep-ph/9803315}; 
I.Antoniadis, N.Arkani-Hamed, S.Dimopoulos and G.Dvali, 
``New Dimensions at a millimeter to a fermi and superstrings at a TeV '', 
Phys. Lett. B {\bf 436} (1998) 257, {\tt hep-ph/9804398}.}
-\ref\HDDi{N.Arkani-Hamed, S.Dimopoulos and G.Dvali, 
``Phenomenology, astrophysics and cosmology of theories with sub-millimeter dimensions 
and TeV scale quantum gravity'', 
Phys. Rev. D {\bf 59} (1999) 086004-1, {\tt hep-ph/9807344}.}
$4 + D$ dimensional theories, where the $D$ extra dimensions are spacelike  
ones of the size $L_{(D)}$, are discussed in order to solve 
the hierarchy problem without using low-energy supersymmetry or technicolor. 
The Planck scale, $M_{Pl(4)}$, of 
the four dimensional theory is related with the one, $M_{Pl(4 + D)}$, of the
$4 + D$ dimensional theories 
as ${M_{Pl(4)}}^2 \sim {L_{(D)}}^D {M_{Pl(4 + D)}}^{D + 2}$. 
For $M_{Pl(4 + D)} \sim 1$TeV 
$L_{(D)}$ is of the order $10^{{30 \over D} - 17}$cm,  
and then $L_{(2)}$ comes out to be a submillimeter for $D = 2$.
When the number $D$ of the extra dimensions increases, the corresponding scale, $L_{(D)}$, 
becomes smaller. So for $D \geq 2$ 
the gravitational force is allowed to feel these large extra dimensions 
without  any conflict to experiments \ADD{},  
but other standard model fields are not \HDDi.
Consequently, our universe is a four dimensional wall in the $4+D$ dimensional space, 
that is a 3-brane, and only gravitons can propagate
in the bulk space, while other standard model fields live in the 3-brane 
and are not able to propagate in the extra dimensions
\ref\RSi{L.Randall and R.Sundrum, ``A Large Mass Hierarchy from a Small Extra Dimension'', 
Phys. Rev. Lett. {\bf 83} (1999) 3370, {\tt hep-ph/9905221}.} 
\ref\RSii{L.Randall and R.Sundrum, ``An Alternative to Compactification'', 
Phys. Rev. Lett. {\bf 83} (1999) 4690, {\tt hep-th/9906064}.}.

On the other hand there is a possibility of the existence of extra timelike
dimensions, which is an interesting problem in its own right 
and has also been discussed 
in various contexts like string theories
\ref\V{C.Vafa, ``Evidence for F-Theory'', Nucl. Phys. {\bf B496} (1996) 403, 
{\tt hep-th/9602022}.}, 
brane theories \ref\CK{M.Chaichian and A.B.Kobakhidze, 
``Mass hierarchy and localization of gravity in extra time'', 
{\tt hep-th/0003269}.}, and so on
\ref\SA{A.D.Sakharov, ``Cosmological transitions with changes in the signature of the metric'', 
Zh. Eksp. Teor. Fix. {\bf 87} (1984) 375 [Sov. Phys. JETP {\bf 60} (1984) 214]; I.Ya.Aref'eva and I.V.Volovich, 
``Kaluza-Klein theories and the signature of space-time'', 
Phys. Lett. B {\bf 164} (1985) 287; 
I.Ya.Aref'eva, B.G.Dragovichi and I.V.Volovich, 
``Extra time-like dimensions lead to a vanishing cosmological constant'', 
Phys. Lett. B {\bf 177} (1986) 357; 
A.Chamblin and R.Emparan, ``Bubbles in Kaluza-Klein theories 
with spacelike or timelike internal dimensions'', 
Phys. Rev. D {\bf 55} (1997) 754, {\tt hep-th/9607236}.}. 
The compactification of timelike curves gives rise to tachyons 
and causes the violation of causality and conserved probability, 
but, as discussed in Ref.\ref\Yn{F.J.Yndur{\' a}in, 
``Disappearance of matter due to causality and probability violations in theories 
with extra timelike dimensions'', 
Phys. Lett. B {\bf 256}(1991)15.}, 
the existence of extra timelike dimensions may not be unphysical 
if the effects of the violation do not conflict with physical observations 
and experiments. 
Similarly Ref.\ref\DGS{
G.Dvali, G.Gabadadze and G.Senjanovi{\' c}, 
``Constraints on extra time dimensions'', {\tt hep-ph/9910207}.} 
studied the Newtonian potential with the tachyon exchange 
in a simplified but unjustified treatment,  
and suggested that the real part of 
the gravitational self-energy of uniform massive body 
with a radius $R < 2 \pi L$, where $L$ 
is the size of extra timelike dimensions, is screened. 
As has been shown rigorously in our previous paper 
\ref\MS{S.Matsuda and S.Seki, 
``Gravitational Stability and Extra Timelike Dimensions'', 
{\tt hep-ph/0007290}, to be published.}, 
for the gravitational self-energy of any spherical massive body 
the correct screening range is actually $R \leq \pi L$ 
contrary to the claim of Ref.\DGS.  

In the following we shall concentrate on the timelike $D$ extra dimensions. 
From the compactification of the $D$ extra dimensions on the timelike circles of a 
radius $L$ we obtain 
tachyonic Kaluza-Klein (KK) modes \Yn .
Let gravitons propagate in the extra dimensions, 
then up to spin factors their propagators are proportional to 
\eqn\pro{
-i {1 \over k_0^2-{\bf k}^2+ {n^2 \over L^2} + i \epsilon},\quad 
n^2 = \sum_{i=1}^D {n_i}^2
}
where $n_i \in {\bf Z}$. Eq.\pro\ is the ordinary massless 
graviton propagator for $n = 0$, 
that is, for $(n_1, \cdots, n_D) = (0, \cdots, 0)$, 
while for $n \neq 0$ it gives the tachyonic KK graviton propagators. 
These tachyonic gravitons normally lead to the imaginary parts of 
the gravitational self-energies for massive bodies,  
thus causing their  gravitational instability, and consequently, 
resulting in  the violation of causality and conserved probability 
in related physical processes \Yn\DGS.

In this paper we let gravitons propagate in the $D$ extra timelike dimensions 
which are  compactified on the timelike circles of the radius $L$. 
Then we investigate the gravitational stability of massive bodies and the screening effect 
of the gravitational force due to the exchange of the KK mode tower of tachyonic gravitons.
In Section 2 we calculate the gravitational self-energies of massive bodies 
with some typical mass densities which are spherically symmetric. 
And in Section 3 we discuss the gravitational stability conditions 
of the spherical massive bodies and 
report on the notable generic relations between the number $D$ of the extra timelike dimensions
and the gravitational stability. 
Section 4 is devoted to the screening effect of the gravitational force, 
whose realization is reported to again depend on the number $D$ of the extra timelike dimensions. 
In Section 5 we discuss the complexity of the Newtonian potential in the $D$ extra times, 
and derive the relation between the Newton constants of the full $4+D$ dimensional spacetime and 
the Newton constant of the ordinary 4 dimensional world. 
In Section 6 we present comments and conclusions. 
We shall give some useful formulas in the appendix.

\newsec{The self energy of spherical bodies}
From Eq.\pro\ we obtain a gravitational potential between two unit mass point
at distance $d$ as
\eqn\pot{
V(d) = - G_N {1 \over d} 
- \sum_{{n=\sqrt{n_1^2 + \cdots + n_D^2}\neq 0 \atop -\infty < n_i < \infty}}
G_N{1\over d}\exp\left( i{n \over L}d \right),
}
in the non-relativistic tree-level approximation, where $G_N$ is the 
Newton constant. The first term is the contribution of ordinary massless 
gravitons and the second is the one of tachyonic KK gravitons, which leads
to the imaginary parts of gravitational self-energies. 
Now, since $\lim_{n \to 0} G_N{1\over d}\exp\left( i{n \over L}d \right) = G_N {1 \over d}$, 
we can rewrite Eq.\pot\ as 
\eqn\poti{
V(d) = - \sum_{{n=\sqrt{n_1^2 + \cdots + n_D^2} \atop -\infty < n_i < \infty}}
G_N{1\over d}\exp\left( i{n \over L}d \right) .
}

We remind the reader at this point that the complex potential used in Refs.\Yn \DGS\  has a wrong 
sign in the phase factor, and that the correct sign of the phase 
together with the correct overall sign of the potential is crucially important in discussing 
the stability of matter from the point of view of its vanishing or explosion \Yn.

Now we consider spherically symmetric bodies of radii $R$ 
with a mass density $\rho(r)$, 
where $r$ is a radial coordinate, and calculate their gravitational self-energies.
We integrate over the opening angle $\beta$ 
between every two mass points of densities $\rho(r)$ and $\rho(l)$ which are apart
from each other at the distance $d = \sqrt{r^2 + l^2 - 2rl\cos\beta}$, and then 
we find the self-energies of spherical bodies to become
\eqn\self{\eqalign{
E_D(R) = &\sum_{{n=\sqrt{n_1^2 + \cdots + n_D^2}\atop -\infty < n_i < \infty}} 
8i \pi^2 G_N L \int_0^R dr \int_0^r dl \rho(r)\rho(l) {rl \over n} f(n, r, l) , \cr
&f(n,r,l) \equiv \exp\left({in \over L}(r + l)\right) 
- \exp\left({in \over L}(r - l)\right). \cr
}}
In order to calculate the summations we assume the discrete numbers $\{n_i\}$ 
as continuous parameters $\{x_i\}$ and replace the infinite summations 
in Eq.\self\ with the integrations over those parameters. 
This replacement is a fairly good approximation 
and will become exact for $r, l \ll L$,
that is, when the size of spherical bodies, $R$, is sufficiently smaller than 
the one of the extra dimensions, $L$.
Then we obtain    
\eqnn\selfx 
$$\eqalignno{
E_D(R) &\approx 8i\pi^2 G_N L \int_0^R dr \int_0^r dl \rho(r)\rho(l) rl 
\int_{-\infty}^{\infty}dx_1 \cdots \int_{-\infty}^{\infty}dx_D 
{f(\sqrt{x_1^2 + \cdots + x_D^2}, r, l) \over \sqrt{x_1^2 + \cdots + x_D^2}} \cr
&= 8i\pi^2 G_N L \int_0^R dr \int_0^r dl \rho(r)\rho(l) rl 
\int d\Omega_{D-1} \int_0^{\infty}dt\ t^{D - 2} f(t,r,l) \cr
&= 8i\pi^2 G_N L S_D
\int_0^R dr \int_0^r dl \rho(r)\rho(l) rl 
\int_0^\infty dt\ t^{D - 2} f(t,r,l) , &\selfx
}$$
where $\int d\Omega_{D-1}$ is the volume of $D-1$ dimensional unit sphere, 
which is $S_D={2\pi^{D/2} \over \Gamma(D/2)}$. 
At this point, by transforming back the integration over  $t$ in the last line of Eq.\selfx\  
to the infinite summation over  discrete non-negative integers,  
the present approximate expression of the self-energies will become more accurate 
and come close to the starting formula, Eq.\self\ .
We then obtain from Eq.\selfx\ 
\eqnn\selfxx 
$$\eqalignno{
E_D(R) &\approx 8i\pi^2 G_N L S_D
\int_0^R dr \int_0^r dl \rho(r)\rho(l) rl
\left[\sum_{n=1}^\infty {1 \over n^{2-D}}f(n, r, l) 
+ {1 \over 2} \lim_{n \to 0} {1 \over n^{2-D}}f(n, r, l)\right] \cr
&= 8i\pi^2 G_N L S_D
\int_0^R dr \int_0^r dl \rho(r)\rho(l) rl
\left[\sum_{n=1}^\infty {1 \over n^{2-D}}f(n, r, l) + i\delta_{D,1}{l \over L}\right]. &\selfxx
}$$
For $D=1$ this equation reproduces the exact result of the gravitational 
self-energy of the spherical body \MS.
The first term in the kernel of Eq.\selfxx\ is the contribution 
of the tachyonic KK gravitons, while the second one is that  
of the ordinary massless gravitons.
For $D \geq 2$ in this approximation the effect of the massless graviton exchange 
is condensed to the first term in the kernel of Eq.\selfxx .

In order to be precise and concrete in our following arguments, 
we next set $\rho(r)$ to some typical densities and 
calculate the gravitational self-energies of the corresponding massive 
spherical bodies as follows.

\subsec{The $\rho(r) = C_0$ (constant) case}
We first set the density of the spherical body as $\rho(r) = C_0$, 
where $C_0$ is a positive constant. This is a simple and normal situation. 
Substituting this density into Eq.\selfxx , we obtain
\eqnn\selfo
$$\eqalignno{
E_D(R) = &8i\pi^2 G_N L S_DC_0^2
\Bigg[\sum_{n=1}^\infty 
\biggl\{ -{iLR^3 \over 3n^{3-D}} - {L^2R^2 \over 2n^{4-D}}
- {L^2R^2 \over 2n^{4-D}}\exp\left( i{2R \over L}n \right) \cr
&- {iL^3R \over n^{5-D}}\exp\left( i{2R \over L}n \right) 
+ {L^4 \over 2n^{6-D}}\exp\left( i{2R \over L}n \right)
- {L^4 \over 2n^{6-D}}\biggr\} + \delta_{D,1}{iR^5 \over 15L}\Bigg]. &\selfo
}$$
So the real part of the self-energy becomes
\eqn\selfore{\eqalign{
\Re E_D(R) = 8\pi^2 G_N S_DC_0^2 L^3
\Bigg[&{R^3 \over 3L}\zeta (3-D) 
+ {R^2 \over 2}\sum_{n=1}^\infty {\sin{2R \over L}n \over n^{4-D}} 
+ LR\sum_{n=1}^\infty {\cos{2R \over L}n \over n^{5-D}} \cr
&- {L^2 \over 2}\sum_{n=1}^\infty {\sin{2R \over L}n \over n^{6-D}} 
- \delta_{D,1}{R^5 \over 15L^3}\Bigg] ,
}}
and the imaginary part is
\eqn\selfoim{\eqalign{
\Im E_D(R) = 8\pi^2 G_N S_DC_0^2 L^3
\Bigg[& - {R^2 \over 2}\zeta (4-D) - {L^2 \over 2}\zeta (6-D) 
- {R^2 \over 2}\sum_{n=1}^\infty {\cos{2R \over L}n \over n^{4-D}}\cr
&+ LR\sum_{n=1}^\infty {\sin{2R \over L}n \over n^{5-D}}
+ {L^2 \over 2}\sum_{n=1}^\infty {\cos{2R \over L}n \over n^{6-D}}\Bigg] ,
}}
where $\zeta$ is the zeta-function.

\subsec{The $\rho(r) = {C_1 \over r}$ case}
Next we consider the case of $\rho(r)={C_1 \over r}$, where $C_1$ is 
a positive constant. 
This is also a physically normal setup. 
It is known that this produces 
interesting results especially for $D=1$ \MS. Using this density we can calculate Eq.\selfxx\ as 
\eqn\selfi{\eqalign{
E_D(R) = &8i\pi^2 G_N L S_DC_1^2
\Bigg[\sum_{n=1}^\infty 
\biggl\{ -{iLR \over n^{3-D}} - {L^2 \over 2n^{4-D}}\exp\left( i{2R \over L}n \right) \cr
&+ {2L^2\over n^{4-D}}\exp\left( i{R \over L}n \right) - {3L^2 \over 2n^{4-D}}
\biggr\} + i \delta_{D,1} {R^3 \over 6L}\Bigg].
}}
From this equation we obtain the real part of the self-energy as 
\eqn\selfire{
\Re E_D(R) = 8\pi^2 G_N S_DC_1^2 L^3
\Bigg[{R \over L}\zeta (3-D) 
+ {1 \over 2}\sum_{n=1}^\infty {\sin{2R \over L}n \over n^{4-D}} 
- 2\sum_{n=1}^\infty {\sin{R \over L}n \over n^{4-D}} - \delta_{D,1}{R^3 \over 6L^3}\Bigg] ,
}
and the imaginary part as 
\eqn\selfiim{
\Im E_D(R) = 8\pi^2 G_N S_DC_1^2 L^3
\left[ - {3 \over 2}\zeta (4-D)  
- {1 \over 2}\sum_{n=1}^\infty {\cos{2R \over L}n \over n^{4-D}} 
+ 2\sum_{n=1}^\infty {\cos{R \over L}n \over n^{4-D}}\right] .
}

\subsec{The $\rho(r) = {C_2 \over r^2}$ case}
At last we let $\rho(r) = {C_2 \over r^2}$, where $C_2$ is a positive constant. 
This is a setup singular at the origin $r=0$.
From Eq.\selfxx\ the self energy 
of the spherical body is 
\eqn\selfii{\eqalign{
E_D(R) = &8i\pi^2 G_N L S_DC_2^2
\Bigg[\sum_{n=1}^\infty {2 \over n^{2-D}}
\bigg\{ \sum_{q=0}^\infty \sum_{k=q}^\infty {(-1)^q \over (2k + 1)^2 (2q)!}
\left({nR \over L}\right)^{2q}\exp \left( i{R \over L}n \right) \cr
&+ i\sum_{q=0}^\infty \sum_{k=q+1}^\infty {(-1)^{q+1} \over (2k + 1)^2 (2q+1)!}
\left({nR \over L}\right)^{2q+1}\exp \left( i{R \over L}n \right)
- \sum_{k=0}^\infty {1 \over (2k+1)^2}
\bigg\} \cr
&+ i\delta_{D,1}{R \over L}\Bigg] .
}}
So the real part of the self-energy is
\eqn\selfiire{\eqalign{
\Re E_D(R) = 8\pi^2 G_N S_DC_2^2 L
\Bigg[2\sum_{q=0}^\infty \bigg\{
\sum_{k=q}^\infty {(-1)^{q+1} \over (2k+1)^2 (2q)!}
\left({R \over L}\right)^{2q}\sum_{n=1}^\infty 
{\sin {R \over L}n \over n^{2-D-2q}} \cr
+ \sum_{k=q+1}^\infty {(-1)^{q} \over (2k+1)^2 (2q+1)!} 
\left({R \over L}\right)^{2q+1}
\sum_{n=1}^\infty {\cos {R \over L}n \over n^{1-D-2q}}\bigg\} 
- \delta_{D,1}{R \over L}\Bigg] ,
}}
and the imaginary part is
\eqn\selfiiim{\eqalign{
\Im E_D(R) = 16\pi^2 G_N S_DC_2^2 L
\Bigg[ \sum_{q=0}^\infty \bigg\{ 
\sum_{k=q}^\infty {(-1)^q \over (2k+1)^2 (2q)!} 
\left({R \over L}\right)^{2q}\sum_{n=1}^\infty 
{\cos {R \over L}n \over n^{2-D-2q}} \cr
+ \sum_{k=q+1}^\infty {(-1)^{q} \over (2k+1)^2 (2q+1)!} 
\left({R \over L}\right)^{2q+1}
\sum_{n=1}^\infty {\sin {R \over L}n \over n^{1-D-2q}}\bigg\} 
- \sum_{k=0}^\infty {1 \over (2k+1)^2} \zeta (2 - D)\Bigg] .
}}

\newsec{The gravitational stability}
As we have seen in the previous section, with the tachyon exchange potential Eq.\pot\ 
the gravitational self-energies of massive spherical bodies normally have 
imaginary parts, which inevitably lead to gravitational instabilities. 
On the other hand, as we have shown in our previous paper \MS\ for the case of $D=1$,   
the spherical bodies which have just right mass densities with right 
discrete values of the radius
such that the imaginary parts of the self-energies vanish, become stable.  

In the following we shall discuss the gravitational stability of massive bodies with higher 
$D$ extra timelike dimensions, $D\geq 1$.

\subsec{The $\rho(r) = C_0$ case}
We first consider the $\rho(r) = C_0$ case. 
For $D=1$ we can calculate the imaginary part of the self-energy 
from Eq.\selfxx\ as
\eqn\oimi{
\Im E_1(R) = -8\pi^2 G_N S_1 C_0^2 L \int_0^R dr \int_0^r dl\ rl \log\left|{\sin{r+l \over 2L} 
\over \sin{r-l \over 2L}}\right|
}
and its numerical result is presented in our previous paper \MS, which says 
that, since $\Im E_1(R)$ does not vanish at any $R$, 
there is no stable radius of the spherical body. 

Using Eqs.\selfoim , (A.4), (A.5) and (A.8),  
for $D=2$ the imaginary part of the self-energy 
at $R$ in the region $0 < R < \pi L$ becomes 
\eqn\oimii{
\Im E_2(R) = -8\pi^2 G_N S_2 C_0^2 L^3\left({\pi\over 6}{ R^3\over L}\right), 
}
and it is proportional to $\sim L^2 R^3$.
 
Using Eqs.\selfoim , (A.6), (A.7) and (A.10), 
for $D=3$ we obtain the imaginary part of the self-energy 
at $0 < R < \pi L$ as
\eqn\oimiii{
\Im E_3(R) = -8\pi^2 G_N S_3
C_0^2 L^3 \left[{1 \over 2}R^2 \left(\zeta (1) - 1 + \log{2R \over L}\right)
+ {\cal O}(R^4)\right], 
}
which diverges since the zeta-function $\zeta(z)$ has a pole at $z = 1$. 
So the spherical body is unstable for $D=3$. 

For $D=4$, using Eqs.\selfoim , (A.8), (A.9) and (A.12), 
the imaginary part of the self-energy 
at $k\pi L < R < (k+1)\pi L\ (k\in \{{\bf N}, 0\})$ 
turns out to be a step function given as 
\eqnn\oimiv
$$\eqalignno{
\Im E_4(R) &= -8\pi^2 G_N S_4 C_0^2 L^3
\left({R^2 \over 2}\zeta(0) + {L^2 \over 2}\zeta(2) 
+ {1 \over 4}R^2 - {\pi^2 \over 12}L^2 -{\pi^2\over 2}k(k+1)L^2  \right) \cr
  &= -8\pi^2 G_N S_4 C_0^2 L^3
\left(
-{\pi^2\over 2}k(k+1)L^2  \right), &\oimiv
}$$
where we have used 
$\zeta(0) = -{1 \over 2}$ and $\zeta(2) = {\pi^2 \over 6}$. 
Thus the spherical bodies with any radii $0 < R < \pi L\ (k=0)$ are stable, while 
at $R>\pi L \  (k\ge 1)$ they are unstable. 

Using Eqs.\selfoim , (A.8), (A.9) and (A.12), 
for $D=5$ we calculate the imaginary part of the self-energy 
at $0 < R < \pi L$ as
\eqn\oimv{
\Im E_5(R) = -8\pi^2 G_N S_5 C_0^2 L^3
\left[{L^2 \over 2}\left(\zeta (1) - {5 \over 4} + \log {2R \over L}\right)
 + {\cal O}(R^6)\right], 
}
and this equation diverges for the same reason as for $D=3$.

For $D=6$, using Eqs.\selfoim , (A.12) and (A.15), 
the imaginary part of the self-energy at all $R$ becomes
\eqn\oimvi{
\Im E_6(R) = -8\pi^2 G_N S_6 C_0^2 L^3
\left({R^2 \over 2}\zeta(-2) + {L^2 \over 2}\zeta(0) + {L^2 \over 4}\right)
= 0,
}
where $\zeta(-2) = 0$ and $\zeta(0) = -{1 \over 2}$. 
Eq.\oimvi\ implies that the spherical bodies can be stable. 

For $D = 2s + 6\ (s \in {\bf N})$ Eq.\selfoim\ is
\eqn\oimvii{\eqalign{
\Im E_{2s+6}(R) = 8\pi^2 G_N S_{2s+6}
C_0^2 L^3 \Bigg( {R^2 \over 2}\zeta (-2-2s) + {L^2 \over 2}\zeta (-2s) 
+ {R^2 \over 2}\sum_{n=1}^\infty {\cos{2R \over L}n \over n^{-2-2s}} \cr
- LR\sum_{n=1}^\infty {\sin{2R \over L}n \over n^{-1-2s}}
- {L^2 \over 2}\sum_{n=1}^\infty {\cos{2R \over L}n \over n^{-2s}}\Bigg) ,
}}
and, substituting Eq.(A.15) into this equation, it becomes
\eqn\oimviii{
\Im E_{2s+6}(R) = 8\pi^2 G_N S_{2s+6}
C_0^2 L^3 \left( {R^2 \over 2}\zeta (-2-2s) + {L^2 \over 2}\zeta (-2s)\right).
}
Since $\zeta(-2t) = 0\ (t \in {\bf N})$, from Eq.\oimviii\ 
we obtain the imaginary part of 
the self-energy at all $R$ as 
\eqn\oimix{
\Im E_{2s+6}(R)= 0, 
}
and then the spherical bodies are stable.

\subsec{The $\rho(r) = {C_1 \over r}$ case}
Next we consider the $\rho(r) = {C_1 \over r}$ case. For $D=1$ we know 
from our previous paper \MS\ 
that, since the imaginary part of the self-energy has a periodicity of $R$ 
with a pitch $2\pi L$ expressed as 
\eqn\iimi{
\Im E_1(2\pi L k + c)= - 8 \pi^2 G_N S_1 C_1^2 L \int_0^c dr \int_0^r dl\  
\log {\left| \sin{r + l \over 2L}\right| \over \sin{r - l \over 2L}}, \quad
k \in \{{\bf N}, 0\}, \quad 0 \leq c < 2\pi L, 
}
the spherical bodies of the radii $R=2\pi L k$ become stable. 
More generally for all $D$ Eq.\selfiim\ also has the same periodicity of $R$ 
with a pitch $2\pi L$, so 
we substitute $R=2\pi L k$ into Eq.\selfiim\ and calculate it as
\eqn\iimixi{
\Im E_D(2\pi L k)=8\pi^2G_NS_D
C_1^2 L^3 \left[-{3 \over 2}\zeta(4 - D) - {1 \over 2}\zeta(4-D)
+ 2 \zeta(4-D)\right]=0.
}
Remarkably enough,  the spherical bodies can be stable at the radii $R=2\pi L k$ 
for any number $D$ of  the extra timelike dimensions.

For $D=2$ from Eqs.\selfiim\ and (A.8) we obtain the imaginary part of the 
self-energy at $0 < R < \pi L$ as 
\eqn\iimii{
\Im E_2(R) = -8\pi^2 G_N S_2 C_1^2 L^3\left({\pi\over 2}{R\over L}\right), 
}
where we have used $\zeta(2) = {\pi^2 \over 6}$. Since Eq.\iimii\ is not zero, 
these spherical bodies are unstable.

For $D=3$, using Eq.(A.10), at $0 < R < \pi L$ Eq.\selfiim\ becomes
\eqn\iimiii{
\Im E_3(R) = 8\pi^2 G_N S_3 C_1^2 L^3 
\left[{1 \over 2}\log 2 - {3 \over 2}\log{R \over L} 
- {1 \over 480}\left({R \over L}\right)^4 - \cdots - {3 \over 2}\zeta(1)\right],
}
where $\zeta(1)$ is a single pole, then the spherical bodies is not stable.

For $D=4$ from Eqs.\selfiim\ and (A.12) we obtain the imaginary part of the 
self-energies at all $R$ as 
\eqn\iimiv{
\Im E_4(R) = 8\pi^2 G_N S_4 C_1^2 L^3
\left(-{3 \over 4} - {3 \over 2}\zeta(0)\right)=0, 
}
where $\zeta(0)=-{1 \over 2}$. Then from Eq.\iimiv\ the spherical bodies 
become stable. 

For $D=5$, using Eqs.\selfiim\ and 
(A.13), we calculate the imaginary part of the self-energy 
at $0 < R < \pi L$ as
\eqn\iimv{
\Im E_5(R) = 8\pi^2 G_N S_5 C_1^2 L^3
\left(-{15 \over 8}{L^2 \over R^2} - {1 \over 128}{R^2 \over L^2} 
- \cdots \right).
}

For $D = 2s + 4\ (s \in {\bf N})$ Eq.\selfiim\ is
\eqn\iimvi{
\Im E_{2s+4}(R) = 8\pi^2 G_N S_{2s+4} C_1^2 L^3
\left[ - {3 \over 2}\zeta (-2s)  
- {1 \over 2}\sum_{n=1}^\infty {\cos{2R \over L}n \over n^{2s}} 
+ 2\sum_{n=1}^\infty {\cos{R \over L}n \over n^{2s}}\right], 
}
and, substituting Eq.(A.15) into Eq.\iimvi\ we obtain
\eqn\iimvii{
\Im E_{2s+4}(R) = 8\pi^2 G_N S_{2s+4} 
C_1^2 L^3\left(- {3 \over 2}\zeta (-2s) \right).
}
Since $\zeta(-2s) = 0$, the imaginary part of the self-energy 
at all $R$ becomes
\eqn\iimvii{
\Im E_{2s+4}(R) = 0, 
}
and the spherical bodies then are stable.

\subsec{The $\rho(r) = {C_2 \over r^2}$ case}
Finally we consider the $\rho(r) = {C_2 \over r^2}$ case.
For $D=1$ Eq.\selfiiim\ includes $\zeta(1)$. So $\Im E_1(R)$ becomes singular.
Using Eqs.(A.12) and (A.15), for $D=2$ at all $R$ Eq.\selfiiim\ becomes
\eqn\iiimi{
\Im E_2(R) = 16\pi^2 G_2 S_2 C_2^2 L
\left[\sum_{k=0}^\infty{1 \over (2k+1)^2}\left(-{1\ \over 2}\right)
- \zeta(0)\sum_{k=0}^\infty{1 \over (2k+1)^2}\right] = 0, 
} 
where we have used $\zeta(0)= - {1 \over 2}$. 
We find the spherical bodies to be stable from Eq.\iiimi . 

And also for $D = 2s + 2\ (s \in {\bf N})$ 
from Eqs.\selfiiim\ and (A.15) the imaginary part of the self-energy 
at all $R$ becomes
\eqn\iiimii{
\Im E_{2s + 2}(R) = 0, 
}
and then the spherical bodies are stable.

\newsec{The screening effect}

Now we consider the real part of the gravitational self-energy and find that 
for a certain number $D$ of the extra timelike dimensions at a certain region of $R$ 
the gravitational self-energy is screened.

\subsec{The $\rho(r) = C_0$ case}
At first we discuss the $\rho(r) = C_0$ case. For $D=1$, 
as previously shown in Ref.\MS , 
from Eq.\selfore\ the real part of the self-energy becomes at $0\leq c < \pi L$
\eqn\orei{\eqalign{
\Re E_1(2\pi L k + c) =& -{8 \over 45}G_N S_1 C_0^2 L^2 \pi^4
(2k+1)k[30c^3 + 15(8k-1)\pi Lc^2 \cr
&+ 60k(2k-1)(\pi L)^2c + (48k^3 - 24k^2 + 2k -1)(\pi L)^3 ],
}}
where $k \in \{{\bf N}, 0\}$, and at $\pi L \leq c < 2\pi L$
\eqn\oreii{\eqalign{
\Re E_1(2\pi L k + c) =& -{8 \over 45}G_N S_1 C_0^2 L^2 \pi^4
(k+1)(2k+1)[30c^3 + 15(8k-3)\pi Lc^2 \cr
&+ 60k(2k -3)(\pi L)^2c + (48k^3 -72k^2 + 74k + 15)(\pi L)^3 ].
}}
Eqs.\orei\ and \oreii\ imply that the gravitational self-energies of 
the spherical bodies of radii $0 \leq R \leq \pi L$ are screened.

Now we concentrate on the spherical bodies of radii $0 < R < \pi L$.
For $D=2$, substituting Eqs.(A.3), (A.6) and (A.7) into Eq.\selfore, 
we obtain 
\eqn\oreiii{
\Re E_2(R) = 8\pi^2 G_N S_2 C_0^2 L^3
\left[{R^3 \over L}\left({1 \over 3}\zeta(1) + {1 \over 3}\log{2R \over L}
- {7 \over 9}\right) + {\cal O}(R^5) \right].
}
Since Eq.\oreiii\ includes $\zeta(1)$, which is a pole, the real part of 
the self-energy diverges. 

For $D=3$, using Eqs.(A.5), (A.8) and (A.9), Eq.\selfore\ becomes
\eqn\oreiv{
\Re E_3(R) = 8\pi^2 G_N S_3 C_0^2 L^3
\left({R^3 \over 3L}\zeta(0) - {1 \over 4}\pi R^2 + {1 \over 6}{R^3 \over L}\right)
= 8\pi^2 G_N S_3 C_0^2 L^3 \left(-{\pi \over 4}R^2 \right),
} 
where $\zeta(0)$ is $-{1 \over 2}$.

For $D=4$ from Eqs.(A.7), (A.10) and (A.11) we calculate Eq.\selfore\ as
\eqnn\orev
$$\eqalignno{
\Re E_4(R) &= 8\pi^2 G_N S_4 C_0^2 L^3
\left({R^3 \over 3L}\zeta(-1) - {3 \over 4}LR + {1 \over 36}{R^3 \over L}
- {1 \over 900}{R^5 \over L^3} + {\cal O}(R^7)\right) \cr
&= 8\pi^2 G_N S_4 C_0^2 L^3
\left(- {3 \over 4}LR - {1 \over 900}{R^5 \over L^3} + {\cal O}(R^7)\right), &\orev
}$$
where $\zeta(-1) = -{1 \over 12}$.

For $D=5$ from Eqs.\selfore , (A.9), (A.12) and (A.15) the real part of 
the self-energy becomes
\eqnn\orevi
$$\eqalignno{
\Re E_5(R) &= 8\pi^2 G_N S_5 C_0^2 L^3
\left[{R^3 \over 3L}\zeta(-2) - {1 \over 2}LR 
- {L^2 \over 4}\left(\pi - {2R \over L}\right)\right] \cr
&= 8\pi^2 G_N S_5 C_0^2 L^3
\left(-{\pi \over 4}L^2\right) &\orevi
}$$
and it is a negative constant independent of $R$. 

For $D=6$, substituting Eqs.(A.11), (A.13) and (A.14) into Eq.\selfore , 
we obtain
\eqnn\orevii
$$\eqalignno{
\Re E_6(R) &= 8\pi^2 G_N S_6 C_0^2 L^3
\left({R^3 \over 3L}\zeta(-3) - {5 \over 8}{L^3 \over R} 
- {1 \over 360}{R^3 \over L} + \cdots \right) \cr
&= 8\pi^2 G_N S_6 C_0^2 L^3 
\left(- {5 \over 8}{L^3 \over R} + {\cal O}(R^5)\right), &\orevii
}$$
where $\zeta(-3)={1\over 120}$ has been used. 
The real part of the self-energy Eq.\orevii\ diverges for $R \to 0$.

For $D=2s + 5\ (s \in {\bf N})$ we have an interesting result 
at all $R$. 
Eq.\selfore\ is then given as 
\eqnn\oreviii
$$\eqalignno{
\Re E_{2s+5}(R) 
= 8\pi^2 G_N S_{2s+5} C_0^2 L^3
\Bigg[&{R^3 \over 3L}\zeta (-2-2s) 
+ {R^2 \over 2}\sum_{n=1}^\infty {\sin{2R \over L}n \over n^{-1-2s}} \cr
&+ LR\sum_{n=1}^\infty {\cos{2R \over L}n \over n^{-2s}} 
- {L^2 \over 2}\sum_{n=1}^\infty {\sin{2R \over L}n \over n^{1-2s}} 
\Bigg]. &\oreviii
}$$
Since $\zeta(-2-2s)$ are zero, Eq.\oreviii\ becomes 
by use of Eq.(A.15)
\eqn\oreix{
\Re E_{2s+5}(R) = 0. 
}
The vanishing of the real part of the self-energy, Eq.\oreix, at all $R$ 
leads to the screening effect.

\subsec{The $\rho(r)={C_1 \over r}$ case}
Next we consider the $\rho(r)={C_1 \over r}$ case. For $D=1$ from \selfire\ 
the real part of the self-energy, which is already given in Ref.\MS , is
\eqn\irei{
\Re E_1(2\pi L k + c) = -16\pi^3 G_N S_1 C_1^2 L 
k [3c^2 + 6\pi Lkc + (4k^2 -1)\pi^2L^2]
}
at $0 \leq c < \pi L$ and 
\eqn\ireii{
\Re E_1(2\pi L k + c) = -16\pi^3 G_N S_1 C_1^2 L 
(k+1)[3c^2 + 6(k-1)\pi L c + (4k^2-4k+3)\pi^2L^2]
}
at $\pi L \leq c < 2\pi L$, where $k \in \{{\bf N}, 0\}$. 
So the real part of the gravitational self-energy with $0 < R < \pi L$ is 
screened.

We now pay attention to the spherical bodies of radii $0 < R < \pi L$ 
for $D\geq 2$. 
Substituting Eq.(A.7) into Eq.\selfire , the real part of the self-energy 
for $D=2$ becomes 
\eqn\ireiii{
\Re E_2(R) = 8\pi^2 G_N S_2 C_1^2 L^3
\left[{R \over L}\left(\zeta(1) + \log{R \over 2L} -1\right)
+ {1 \over 36}{R^3 \over L^3} + {\cal O}(R^5)\right], 
}
which can not converge because $\zeta(1)$ is a pole. 

For $D=3$ from Eqs.\selfire\ and (A.15) the real part of the self-energy is
\eqnn\ireiv
$$\eqalignno{
\Re E_3(R) &= 8\pi^2 G_N S_3 C_1^2 L^3
\left[{R \over L}\zeta(0) + {1 \over 4}\left(\pi - {2R \over L}\right) 
- \left(\pi - {R \over L}\right)\right]  \cr
&=8\pi^2 G_N S_3 C_1^2 L^3
\left(-{3 \over 4}\pi\right), &\ireiv
}$$
where $\zeta(0) = -{1 \over 2}$ and it is a negative constant.

For $D=4$ from Eq.(A.11) the real part of the self-energy Eq.\selfire\ becomes
\eqn\irev{
\Re E_4(R) = 8\pi^2 G_N S_4 C_1^2 L^3
\left(-{7 \over 4}{L \over R} - {1 \over 360}{R^3 \over L^3} 
+ {\cal O}(R^5)\right), 
}
where we have used $\zeta(-1) = -{1 \over 12}$.
The result diverges when $R\rightarrow 0$.

For $D = 2s + 3\ (s \in {\bf N})$ we also have a characteristic result
at any $R$. From Eq.\selfire\ we obtain
\eqn\irevi{
\Re E_{2s+3}(R) = 8\pi^2 G_N S_{2s+3} C_1^2 L^3
\left[{R \over L}\zeta (-2s) 
+ {1 \over 2}\sum_{n=1}^\infty {\sin{2R \over L}n \over n^{1-2s}} 
- 2\sum_{n=1}^\infty {\sin{R \over L}n \over n^{1-2s}} \right]. 
}
And, since we know Eq.(A.15) and $\zeta(-2s)=0$ to hold, Eq.\irevi\ becomes
\eqn\irevii{
\Re E_{2s+3}(R) = 0.
}
So Eq.\irevii\ leads to the screening of the gravitational self-energy at all $R$.

\subsec{The $\rho(r)={C_2 \over r^2}$ case}
We consider the $\rho(r)={C_2 \over r^2}$ case.
For $D=1$ at $0 < R < \pi L$ from Eqs.\selfiire , (A.9), (A.12) 
and (A.15) we obtain with the formula 
$\sum_{k=0}^\infty {1 \over (2k+1)^2}={\pi^2\over 8}$ 
\eqnn\iirei
$$\eqalignno{
\Re E_1(R) &= 8\pi^2 G_N S_1 C_2^2 L
\left[2
\left(-{1 \over 2}\pi \sum_{k=1}^\infty {1 \over (2k+1)^2} -{\pi \over 2} + 
{1 \over 2}{R \over L}\right) - {R \over L}\right] \cr
&= 8\pi^2 G_N S_1 C_2^2 L
\left(-{\pi^3\over 8} \right), &\iirei
}$$
which shows that the real part of the self-energy becomes a negative constant.

For $D = 2s + 1\ (s \in {\bf N})$ Eq.\selfiire\ becomes
\eqnn\iireii
$$\eqalignno{
\Re E_{2s+1}(R) =& 
8\pi^2 G_N S_{2s+1} C_2^2 L
\Bigg[2\sum_{q=0}^\infty \Bigg\{
\sum_{k=q}^\infty {(-1)^{q+1} \over (2k+1)^2 (2q)!}
\left({R \over L}\right)^{2q}\sum_{n=1}^\infty 
{\sin {R \over L}n \over n^{1-2s-2q}} \cr
&+ \sum_{k=q+1}^\infty {(-1)^{q} \over (2k+1)^2 (2q+1)!} 
\left({R \over L}\right)^{2q+1}
\sum_{n=1}^\infty {\cos {R \over L}n \over n^{-2s-2q}}\Bigg\} \Bigg]. &\iireii 
}$$
Since from Eq.(A.15) the equalities 
$
\sum_{n=1}^\infty {\sin {R \over L}n \over n^{1-2s-2q}}
= \sum_{n=1}^\infty {\cos {R \over L}n \over n^{-2s-2q}} = 0
$ hold, 
we can calculate the real part of the self-energy \iireii\ at any $R$ as
\eqn\iireiii{
\Re E_{2s+1}(R) = 0.
}
Then the gravitational self-energy is screened at all $R$.

\newsec{The Newton constants in $D$ extra timelike dimensions}
The correlation between the complexity of the Newtonian potential 
and the number $q$ of  extra times has been discussed in Ref.\DGS\   from somewhat 
different perspective. 
It is pointed out there that the Newtonian potential $m^2 V(d)$ between 
two point-like masses $m$ which are localized at a particular time moment $\tau=0$ 
in  $q$ extra times at $d\ll L$ distance apart becomes pure imaginary for odd $q$ like 
$$
m^2 V(d) \sim (i)^q{m^2\over M_{Pl(4+q)}^{2+q}}{1\over d^{1+q}}
$$
for the static case, meaning that the Newtonian potential is screened 
at  $d\ll L$ distances.
Our work is expected to shed some light on this line of investigation as well.

In fact, in this section we shall derive the exact relationship between the Newton constants 
$\hat G_{N(4+D)}$  in the $D$ extra timelike dimensions and
the ordinary 4 dimensional Newton constant $G_N=\hat G_{N(4)}$ with $D=0$, 
which are defined by the attractive force laws between two mass points $m_1$, $m_2$ 
at distance $d$ 
\eqnn\force
$$\eqalignno{
\hat F_{(4+D)}(d) &=-\hat G_{N(4+D)}{m_1m_2\over d^{2+D}},  \cr
\hat F_{(4)}(d) &=-G_N{m_1m_2\over d^2}. &\force
}$$

Now we start with the gravitational potential Eq.\poti\  between two unit mass points. 
As we have done in Section 2, 
we perform the replacement in Eq.\poti\  to transform the infinite summation to the integration as 
\eqnn\gr
$$\eqalignno{
m^2V(d)   &\approx -G_N{m^2\over d}
\int_{-\infty}^{\infty}dx_1 \cdots \int_{-\infty}^{\infty}dx_D 
             \exp\left(i{d\over L}\sqrt{x_1^2+\cdots+x_D^2}\right) \cr
&=  -G_N{m^2\over d}
\int d\Omega_{D-1} \int_0^{\infty}dt\ t^{D - 1}  \exp\left(i{d\over L}t\right)  \cr
&=  -G_N{m^2\over d}
S_D\int_0^\infty dt\ t^{D - 1} \exp\left(i{d\over L}t\right)   , &\gr 
 }$$
where $S_D={2\pi^{D \over 2} \over \Gamma({D \over 2})}$. 
Note that this equation becomes exact when $d\ll L$, that is, for the limit $d/L\rightarrow 0$. 
By deforming the integration contour through the analytic continuation method, 
we finally obtain with the $D$ dimensional timelike volume $V_D=(2\pi L)^D$
\eqnn\gri
$$\eqalignno{
m^2V(d) &=-(i)^D G_N S_D \Gamma(D)L^D {m^2\over d^{1+D}}  \cr
   &=-(i)^D G_N {S_D \Gamma(D)V_D\over (2\pi)^D} {m^2\over d^{1+D}}  \cr
   &=-{\hat G_{N(4+D)}\over 1+D}{m^2\over d^{1+D}}, &\gri
}$$
which implies that the following relationship holds between the Newton constants for the 
$D$ extra timelike dimensions and the Newton constant of our world
\eqn\grii{
 \hat G_{N(4+D)}=(i)^D{4\pi V_D\over S_{(3+D)}}G_N.
}
Thus we find that, since $G_N$ is real, the Newton constant $\hat G_{N(4+D)}$ of 
the full $4+D$ dimensional theories with the $D$ extra times becomes pure imaginary 
for odd $D$, while for even $D$ it is pure real.

\newsec{Conclusions}
We discussed the $D$ extra timelike dimensions and compactified them 
on the circles of the radius $L$. Then the tachyonic Kaluza-Klein modes 
are induced. We let only the gravitons propagate in the $D$ extra timelike 
dimensions. And we calculated the gravitational self-energies of spherical 
bodies of radii $R$ in the fairly good approximation. 
Note that for $D=1$ the results are exact. 
The tachyonic KK gravitons give rise to the imaginary 
parts of the self-energies, which leads to the instability of the spherical 
bodies. In some dimensions the contribution of the ordinary massless 
gravitons to the self-energies are canceled out by the one of the tachyonic 
KK gravitons, and the self-energies of the spherical bodies of certain radii 
with certain mass densities are screened.

We considered the imaginary parts of the self-energies of spherical bodies 
with the three typical and spherically symmetric mass densities and 
discussed their stability. 
At first we set the density $\rho(r) = C_0$. 
From Eqs.\oimvi\ and \oimix\ the imaginary parts of 
the self-energies of the spherical bodies which have any radii $R$ 
vanish for $D = 2s + 4$  $(s \in {\bf N})$ then the spherical bodies 
are stabilized. And also for $D=4$ the imaginary part of the self-energy 
becomes zero at each value $R$ for the range $0 < R < \pi L$ from Eq.\oimiv, 
so the spherical body is stable.
Since for $D = 3$ and $5$ Eqs.\oimiii\ and \oimv\ involve 
$\zeta(1)$, which is a pole, the imaginary parts of the self-energies diverge 
at $0 < R < \pi L$. 
Next we let the density $\rho(r) = {C_1 \over r}$. 
This has the interesting features. Eq.\iimixi\ shows that the spherical bodies 
which have critical radii $R=2\pi L k$ $(k \in \{0, {\bf N}\})$ are stable 
for any dimension $D$. 
And from Eqs.\iimiv\ and \iimvii\  for $D =2s+2$  $(s \in {\bf N})$,  the imaginary parts 
of the self-energies become identically zero at all $R$,  
so the corresponding spherical bodies with any value of the radii $R$ become stable. 
At last we adopted $\rho(r) = {C_2 \over r^2}$ as the density.
For $D=2s$  $(s \in {\bf N})$, from Eqs.\iiimi\ and \iiimii\ the imaginary parts 
of the gravitational self-energies again vanish at all $R$, then
the spherical bodies of any radii $R$ 
are again stable.

And we discussed the screening effects due to the tachyonic KK gravitons 
which are signaled by the vanishing of the real parts of the self-energies. 
When $\rho(r)$ has the constant value $C_0$, 
at the region $0 < R < \pi L$ the gravitational force is 
screened for $D=1$ from Eq.\orei . 
At that region of $R$ from Eq.\oreiii\ the real part of the self-energy 
for $D=2$ diverges because of the pole of $\zeta(1)$ and from Eq.\orevi\ 
the one for $D=5$ becomes a negative constant independent of $R$. 
On the other hand, for $D=2s+5$ $(s \in {\bf N})$, from Eq.\oreix\ we can say that the 
gravitational force is screened at any $R$. 
We next considered the $\rho(r)= {C_1 \over r}$ case.  
Eq.\irevii\ implies that for $D=2s+3$ $(s \in {\bf N})$ 
the gravitational forces are again screened at all  R.
At $0<R<\pi L$ from Eq.\irei\   the real part of the gravitational self-energy for $D=1$ vanishes, 
resulting in the screening of the gravitational force. 
At the same region from Eq.\ireiii\ the real part of the self-energy for 
$D=2$ becomes a pole and from Eq.\ireiv\ the one for $D=3$ is 
a negative constant, so we do not have a screening for these cases. 
Lastly we set $\rho(r) = {C_2 \over r^2}$. 
From Eq.\iirei\ the real part of the self-energy becomes a negative constant  
for $D=1$, resulting in no screening, 
while from Eq.\iireiii\ the gravitational forces are screened 
for $D=2s+1$ $(s \in {\bf N})$ at all R due to the vanishing of 
the real parts of the corresponding self-energies. 

On the last choice of $\rho(r)={C_2 \over r^2}$ we have a comment in order 
for the case $D=1$ of the extra timelike dimension. 
This choice of $\rho(r)$ is very singular and causes divergences 
in the exact formula of the gravitational self-energy for $D=1$ 
presented in our previous paper \MS, 
thus ruining the general statement that for $D=1$ the gravitational force is 
screened for any spherical mass density $\rho (r)$ at $0<R<\pi L$. 
Nevertheless, the claim in the statement remains to be generically valid for any $\rho(r)$ 
of reasonably mild analytic property.

We discussed the remarkable correlation between the number $D$ of the extra times and 
the complexity of the Newtonian potential. 
In particular, we derived the exact relationship between the ordinary Newton constant $G_N$ 
in our 4 dimensions and 
the Newton constant $\hat G_{N(4+D)}$ of the full $4+D$ dimensional spacetime 
with the $D$ extra times.   

In our whole investigations we replaced the infinite summation \self\ 
with the integration \selfx\ .
This is the fairly good approximation in general and the replacement becomes exact 
for $l, r \ll L$, $0 \leq l, r \leq R$, 
or equivalently for $R \ll L$. 
And by transforming the integration back to the infinite 
summation \selfxx\ again, that approximation recovers a reasonable accuracy. 
In fact for $D=1$ Eq.\selfxx\ is the exact result. At the region $R \ll L$ 
our consideration about the gravitational stabilities and screening effects 
can be regarded as precise, but at the other region it may be less reliable. 
This implies that at least we could say that 
if the extra timelike dimensions exist and the spherically 
symmetric body with a certain density is stable, the scale of the extra 
dimensions must be sufficiently larger than the one of the spherical body.

We studied the relations among the extra timelike dimensions, the 
stability of the spherical bodies and the screening effects of 
the gravitational force.
These relations may give us some useful means for determining the size of 
particles or universes and for constructing some theories 
which include gravity.

\bigbreak\bigskip
\bigskip\centerline{{\bf Acknowledgments}}
This work is supported in part by the Grant-in-Aid 
for Scientific Research on Priority Area 707 
``Supersymmetry and Unified Theory of Elementary Particles", 
Japan Ministry of Education. S. M. is also funded partially 
by the Grant-in-Aid for Scientific Research (C) (2)-10640260, 
while S. S. is supported in part by JSPS Research Fellowship 
for Young Scientists.

\appendix{A}{Infinite summations}
The following infinite summations are known at $0 < x < 2\pi$ to hold:
\eqnn\ai \eqnn\aii \eqnn\aiii \eqnn\aiv \eqnn\av \eqnn\avi \eqnn\avii 
\eqnn\aviii \eqnn\aix \eqnn\ax \eqnn\axi \eqnn\axii \eqnn\axiii \eqnn\axiv
\eqnn\axv
$$
\eqalignno{
&\sum_{n = 0}^\infty {\sin nx \over n^5} 
= {\pi^4 \over 90}x - {\pi^2 \over 36}x^3 + {\pi \over 48}x^4 - {1 \over 240}x^5 , &\ai
\cr
&\sum_{n = 0}^\infty {\cos nx \over n^5} 
= \zeta(5) - {\zeta(3) \over 2}x^2 - {1 \over 24}x^4\log x + {25 \over 288}x^4 + {1 \over 8640}x^6 + \cdots, &\aii
\cr
&\sum_{n = 0}^\infty {\sin nx \over n^4} 
= \zeta(3)x + {1 \over 6}x^3\log x - {11 \over 36}x^3 - {1 \over 1440}x^5 - \cdots, &\aiii
\cr
&\sum_{n = 0}^\infty {\cos nx \over n^4} 
={\pi^4 \over 90} - {\pi^2 \over 12}x^2 + {\pi \over 12}x^3 - {1 \over 48}x^4 , &\aiv
\cr
&\sum_{n = 0}^\infty {\sin nx \over n^3} 
= {\pi^2 \over 6}x - {\pi \over 4}x^2 + {1 \over 12}x^3 , &\av
\cr
&\sum_{n = 0}^\infty {\cos nx \over n^3} 
= \zeta (3) + {x^2 \over 2}\log 2 + \int_0^x (x-t)\log\left(\sin{t \over 2}\right)dt
\cr
&\phantom{\sum_{n = 0}^\infty {\cos nx \over n^3}}
= \zeta (3) + {1 \over 2}x^2\log x - {3 \over 4}x^2 - {1 \over 288}x^4 - {1 \over 86400}x^6 - \cdots, &\avi
\cr
&\sum_{n = 0}^\infty {\sin nx \over n^2} 
= -x\log 2 - \int_0^x \log\left(\sin{t \over 2}\right)dt 
\cr
&\phantom{\sum_{n = 0}^\infty {\sin nx \over n^2}} 
= -x\log x + x + {1 \over 72}x^3 + {1 \over 14400}x^5 + \cdots, &\avii
\cr
&\sum_{n = 0}^\infty {\cos nx \over n^2} 
= {\pi^2 \over 6} - {\pi \over 2} x + {1 \over 4}x^2 , &\aviii
\cr
&\sum_{n = 0}^\infty {\sin nx \over n} 
= {\pi \over 2} - {1 \over 2}x, &\aix
\cr
&\sum_{n = 0}^\infty {\cos nx \over n} 
= -\log \left( 2 \sin {x \over 2} \right) 
= -\log x + {1 \over 24}x^2 + {1 \over 2880}x^4 + \cdots , &\ax
\cr
&\sum_{n = 0}^\infty \sin nx 
= {1 \over x} - {x \over 12} - {x^3 \over 720} + \cdots , &\axi
\cr
&\sum_{n = 0}^\infty \cos nx 
= - {1 \over 2}, &\axii
\cr
&\sum_{n = 0}^\infty n \cos nx 
= - {1 \over x^2} - {1 \over 12} - {1 \over 240}x^2 + \cdots , &\axiii
\cr
&\sum_{n = 0}^\infty n^2 \sin nx 
= -{2 \over x^3} + {x \over 120} + \cdots , &\axiv
\cr
&\sum_{n = 0}^\infty n^{2p}\cos nx 
= \sum_{n = 0}^\infty n^{2p-1}\sin nx = 0,\quad p \in {\bf N}. &\axv
}$$
Eqs.(A.11)-(A.15) can  be obtained by use of the equality (15) in page 30 of Ref.
\ref\Er{{\it Higher Transcendental Functions, Vol.1, Bateman Manuscript Project}, Ed. A. Erd{\' e}lyi, 
McGraw-Hill Book Company, Inc. (1953).}.

\listrefs
\bye